\documentclass[preprint,authoryear,12pt,3p,a4paper]{elsarticle1}

\usepackage[latin1]{inputenc}
\usepackage{graphicx}
\usepackage{amsmath}
\usepackage{verbatim} 

\newcommand{\e}[1]{\mathrm{e}^{#1}}
\newcommand{\D}{\mathrm{D}}
\newcommand{\rmd}{\mathrm{d}}
\newcommand{\m}{\mathrm{m}}
\newcommand{\gd}{{\ast}}
\newcommand{\PD}{P_\mathrm{D}}

\journal{$ $}

\begin{document}

\title{Calibration of structural and reduced--form recovery models}

\author{Alexander F. R. Koivusalo}
\address{Danske Capital, Copenhagen, Denmark}
\address{Faculty of Physics, University of Duisburg-Essen, Germany}

\author{Rudi Schäfer}
\ead{rudi.schaefer@uni-duisburg-essen.de}
\address{Faculty of Physics, University of Duisburg-Essen, Germany}

\date{\today}

\begin{abstract}
In recent years research on credit risk modelling has mainly focused on default probabilities.
Recovery rates are usually modelled independently, quite often they are even assumed constant.
Then, however, the structural connection between recovery rates and default probabilities is lost and the tails of the loss distribution can be underestimated considerably. 
The problem of underestimating tail losses becomes even more severe, when calibration issues are taken into account.
To demonstrate this we choose a Merton--type structural model as our reference system. Diffusion and jump--diffusion are considered as underlying processes.
We run Monte Carlo simulations of this model and calibrate different recovery models to the simulation data. 
For simplicity, we take the default probabilities directly from the simulation data.
We compare a reduced--form model for recoveries with a constant recovery approach. In addition, we consider a functional dependence between recovery rates and default probabilities. This dependence can be derived analytically for the diffusion case.
We find that the constant recovery approach drastically and systematically underestimates the tail of the loss distribution.
The reduced--form recovery model shows better results, when all simulation data is used for calibration.
However, if we restrict the simulation data used for calibration, the results for the reduced--form model deteriorate.
We find the most reliable and stable results, when we make use of the functional dependence between recovery rates and default probabilities.
\end{abstract}


%\keywords{Credit risk, Loss distribution, Reduced--form models, Structural models, Value at Risk, Expected Tail Loss, Stochastic processes }

\maketitle

\vfill

\newpage

\section{Introduction}  \label{sec1}

To correctly assess and control exposure to credit risk is a vital question for financial institutions. It is also a crucial problem for banking regulation.
There are two conceptually different approaches to credit risk modelling: structural and reduced--form approaches.
The structural models go back to \cite{black73} and \cite{Merton74}.
The Merton model is based on the assumption that a company has a certain amount of zero--coupon debt which becomes due at a fixed maturity date. The market value of the company is modelled as a stochastic process. A possible default and the associated recovery rate are determined directly from this market value at maturity.
In the reduced--form approach default probabilities and recovery rates are described independently by stochastic models. The aim is to describe the dependence of these quantities on common covariates or risk factors.
For some well known reduced--form model approaches see, e.g., \cite{JarrowTurnbull95}, \cite{JarrowLandoTurnbull97}, 
\cite{DuffieSingleton99}, \cite{HullWhite2000} and  \cite{Schoenbucher2003}.
First Passage Models constitute a third approach which is usually regarded as structural, but is better described as a mixed or pseudo--structural approach. First Passage Models were first introduced by \cite{BlackCox1976}.
As in the Merton model, the market value of a company is modelled as a stochastic process.
Default occurs as soon as the market value falls below a certain threshold.
In contrast to the Merton model, default can occur at any time.
In this approach, the recovery rate is not determined by the underlying process for the market value.
Instead, recovery rates are modelled independently, for example, by a reduced--form approach (see e.g., \cite{AsvanuntStaal09a, AsvanuntStaal09b}).  
In some cases recovery rates are even assumed constant, for instance, in \cite{Giesecke2004}. 

The independent modelling of default and recovery rates can lead to a serious underestimation of large losses. This situation is particularly troublesome when considering calibration to limited default and recovery data.
One way to address the scarcity issue of default and recovery data is to calibrate to market data of credit contracts and derivatives instead.
This allows for a self--consistent description of market prices, but may misrepresent the actual credit risk. As witnessed in the recent financial crisis, the market as a whole may be completely wrong in its pricing consensus.
Therefore it is crucial for the risk assessment of a credit portfolio to take into account historical data of defaults and recoveries.

The aim of our study is to discuss general calibration issues that arise from independent modelling of default and recovery rates.
To this end we use a structural model as our reference system. While the original Merton model contains only a diffusion process, we also consider a jump--diffusion process for the evolution of the market value.
For related works on jump--diffusion processes in credit risk modelling see, e.g., \cite{zhou01},
\cite{schaefer07a} and \cite{KieselScherer2007}.
We run Monte Carlo simulations of this model and calibrate different recovery models to the simulation data. 
For simplicity, we take the default probabilities directly from the simulation data.
We compare a reduced--form model for recoveries with a constant recovery approach. In addition we consider a functional dependence between recovery and default rates. This dependence can be derived analytically for the diffusion case.
We find that the constant recovery approach drastically and systematically underestimates the tail of the loss distribution.
The reduced--form recovery model shows better results, when all simulation data is used for calibration.
However, if we restrict the calibration data, the results for the reduced--form model deteriorate.
We find the most reliable and stable results, when we make use of the functional dependence between recovery and default rates.

The paper is organized as follows. In Section~\ref{sec2} we discuss the structural model used as our reference system. In Section~\ref{sec3} we describe the recovery rate models which will be calibrated to the Monte Carlo data. Issues that arise in model calibration are discussed in Section~\ref{sec4}. 
In Section~\ref{sec5} we compare the different recovery models with respect to three central questions: To what extend do the models preserve the dependence of recovery and default rates? How robust are the models with respect to calibration issues? And do they provide accurate estimates for the risk of large portfolio losses? We summarize our findings and conclude in Section~\ref{sec6}.

\section{Structural model as reference system} \label{sec2}

In the Merton model, defaults and recoveries are directly determined by an underlying market value at maturity. Hence, stochastic modelling of the market value $V_k(t)$ of a company allows to assess its credit risk.
The model assumes that a company $k$ has a certain amount of zero--coupon debt with a face value $F_k$. The debt will become due at maturity time $T$. 
The company defaults if the value of its assets at time $T$ is less than the face value, i.e., if $V_k(T)<F_k$.
The recovery rate is then $R_k=V_k(T)/F_k$ and the normalized loss given default is 
\begin{equation}\label{eq:lossgd}
L_k^\gd=1-R_k=\frac{F_k-V_k(T)}{F_k}  \;.
\end{equation}
We denote the loss given default with an asterisk to distinguish it from the loss including non--default events. The individual loss can be expressed as 
\begin{equation}\label{eq:lossk}
L_k=\left( 1-\frac{V_k(T)}{F_k} \right) \, \Theta\left( 1-\frac{V_k(T)}{F_k} \right) \;,
\end{equation}
where $\Theta$ is the Heaviside function.

\subsection{Model setup}
We model the time evolution of the market value of a single company $k$ 
by a stochastic differential equation of the form
\begin{equation} \label{eq:jumpdiff}
\frac{\rmd V_k}{V_k} = \mu \rmd t +  \sqrt{c}\, \sigma \rmd W_{\rm m} + \rmd J_ {\rm m}
 + \sqrt{1-c}\, \sigma \rmd W_k + \rmd J_k \ .
\end{equation}
This is a correlated jump--diffusion process with a deterministic term $\mu\rmd t$ and
a linearly correlated diffusion. The Wiener processes $\rmd W_k$ and 
$\rmd W_{\rm m}$ describe the idiosyncratic and the market fluctuations, respectively.
For simplicity, we choose the drift $\mu$, the volatility $\sigma$ and the correlation coefficient $c$ as constant parameters, which are the same for all companies.
%%%
The jump terms are not contained in Merton's original model; they ensure that the default probability does not go to zero as the time to maturity becomes very short.
We include two jump terms in the stochastic process, $\rmd J_k$ for idiosyncratic jumps and $\rmd J_ {\rm m}$ for jumps which affect the entire market. 
As in \cite{schaefer07a}, the jumps are modelled by a {\it Poisson process} with intensity $\lambda$. We
recall that in such a process the probability function for the event
to occur $n$ times between zero and the time $t$ is given by
\begin{equation}\label{pp}
p_n^{{\rm Poisson}}(t) = 
  \frac{(\lambda t)^n}{n!}\exp\left(-\lambda t\right) \ .
\end{equation}

The size $\Lambda$ of the jump, measured in units of the current asset
value $V(t)$, is a random variable with a distribution which we have
to specify.  Jumps can be positive or negative. The largest possible
negative jump is 100\% of the current asset value. Based on this
information, a possible distribution of the jump size $\Lambda$ is a
shifted lognormal distribution, 
$\Lambda + 1 \sim {\rm LN}(\mu_J,\sigma_J)$, with mean $\mu_J$ and standard deviation
$\sigma_J$.

Without the jump term, the distribution of the asset price $V_k(t)$ is
log--normal. The jumps render the tails of the asset price
distribution fatter. Fat tails are empirically
observed~\cite{mant00}. As this clearly affects the loss distribution,
we find it important to include such jumps.
The parameters of the jump process can be adjusted in order to match the tail behavior of a given empirical time series of the asset value. We use the same parameters for idiosyncratic and market wide jumps.

\subsection{Monte Carlo simulation}

In the Monte Carlo simulations we consider the stochastic process in Equation~(\ref{eq:jumpdiff}) for discrete time increments $\Delta t=T/N$, where the time to maturity $T$ is divided into $N$ steps. 
The market value of company $k$ at maturity is then given by
\begin{equation}\label{eq:VkT}
V_k(T)=V_k(0) \prod\limits_{t=1}^{N} 
\left(1+ \mu \Delta t +  \sqrt{c}\, \sigma \eta_{{\rm m},t} \sqrt{\Delta t} + \rmd J_{{\rm m},t}
 + \sqrt{1-c}\, \sigma \varepsilon_{k,t} \sqrt{\Delta t} + \rmd J_{k,t}\right) \;.
\end{equation}
The random variables $\eta_{{\rm m},t}$ and $\varepsilon_{k,t}$ are independent and are drawn from a normal distribution. 
We consider a homogeneous portfolio of size $K$ with the same parameters for each asset process, and with the same face value, $F_k=F$, and initial market value, $V_k(0)=V_0$.
The simulation is run with an inner loop and an outer loop.
In the inner loop we simulate $K=500$ different realizations of $\varepsilon_{k,t}$ and $\rmd J_{k,t}$ for a single realization of the market terms $\eta_{{\rm m},t}$ and $\rmd J_{{\rm m},t}$ with $t=1,\ldots,N$. 
The inner loop can be interpreted as a homogeneous portfolio of size $K$, or simply as an average over the idiosyncratic part of the process. 
In each run of the inner loop, we calculate the market return $X_{\rm m}$, the number of defaults $N_\D(X_{\rm m})$ and the expected recovery rate $\left<R(X_{\rm m})\right>$.
The market return $X_{\rm m}$ is defined as the average return at maturity,
\begin{equation}
X_{\rm m} = \frac{1}{K}\sum\limits_{k=1}^K \left( \frac{V_k(T)}{V_k(0)}-1 \right)  \;.
\end{equation}
For sufficiently large $K$ the idiosyncratic terms average out and the market return $X_{\rm m}$ is solely defined by the realizations of $\eta_{{\rm m},t}$ and $\rmd J_{{\rm m},t}$. This is why we use the market return as a parameter for the other observables.
The number of defaults $N_\D(X_{\rm m})$ simply counts how many times the condition $V_k(T)<F$ is fulfilled. We can estimate the default probability as 
\begin{equation}\label{eq:pd}
\PD(X_{\rm m})\approx N_\D(X_{\rm m})/K \;.
\end{equation}
We obtain the portfolio loss as the average of individual losses in Equation~(\ref{eq:lossk}),
\begin{equation}\label{eq:ploss}
\left< L(X_{\rm m}) \right> = \frac{1}{K} \sum_{k=1}^{K} L_k \;.
\end{equation}
Using the relation 
\begin{equation}\label{eq:plossrec}
\left< L(X_{\rm m}) \right> = \PD(X_{\rm m}) \left( 1-\left<R(X_{\rm m})\right> \right) 
\end{equation}
we can estimate the expected recovery rate as
\begin{equation}\label{eq:expectedR}
\left< R(X_{\rm m}) \right> 
= 1- \frac{\left<L(X_{\rm m})\right>}{\PD(X_{\rm m})} 
\approx 1- \frac{K \left<L(X_{\rm m})\right>}{N_\D(X_{\rm m})} \;.
\end{equation}
Here, we assume that the number of defaults is strictly non--zero, which is justified for large $K$.
The outer loop runs over $10^6$ realizations of the market terms, where we obtain different values for the market return $X_{\rm m}$ and, consequently, the number of defaults $N_\D(X_{\rm m})$ and the expected recovery rate $\left<R(X_{\rm m})\right>$.

% PARAMETER

The Monte Carlo simulations are carried out for diffusion and jump diffusion. 
In this paper we do not discuss the parameter dependence of the models or aim at calibrating them to a given portfolio. 
Instead, we only present the results for a single set of parameters with economically sensible values.
This suffices to make our main argument about different recovery model approaches.
As correlation coefficient we choose $c=0.5$. The parameters for the diffusion process are $\mu=0.05$ and $\sigma=0.15$. 
The additional parameters for the jump terms are $\lambda=0.005$, $\mu_{\mathrm{j}}=0.4$ and $\sigma_{\mathrm{j}}=0.3$.
The initial market value is set to $V_0=100$, the face value of the zero--coupon bonds is $F=75$ with maturity time $T=1$.

\section{Recovery rate models} \label{sec3}

In our study we want to isolate the influence of the different recovery models on various risk measures. Therefore we will not employ a stochastic description of default probabilities. Instead we   consider the default data from the simulations directly.
We compare three different recovery rate models: a reduced--form model, constant recoveries and structural recoveries.

\subsection{Reduced--form approach}

Reduced--form models are stochastic models for default probabilities and recovery rates. Their aim is to  describe the dependence of these quantities on common (macroeconomic) covariates or risk factors.
In our structural reference model we use a one-factor model with a constant correlation of all companies to the market. Therefore we use the market return, i.e., the average return of all assets at maturity, as a covariate.

Default events are commonly modelled by a 
doubly stochastic Poisson process or {\it Cox process}.  %
As mentioned above, we will not consider a stochastic description of default probabilities in this paper. Instead we consider the default data from the simulations directly. % 

Reduced--form recovery rate models describe a monotonous dependence on covariates. In our case, for instance, we expect a lower recovery rate for lower market returns. Since the natural domain for the recovery rate is the interval $[0,1]$, a generalized linear model is typically used.
In this approach a linear dependence on the covariates is transformed by a link function to the desired domain.
Recovery rate models commonly use either a logit or a probit link function.  
A logit link function is used, eg, in \cite{Schoenbucher2003}, a probit link function is, eg, in \cite{AndersenSidenius2005}.
As pointed out by \cite{Chava2008}, both link functions lead to very similar results. 
In our study we focus on the probit model; then the reduced--form recovery rate model reads
\begin{equation}
\langle R(X_\m)\rangle =\Phi(-\gamma X_\m - \delta) \label{recovery}
\end{equation}
with the parameters $\gamma$ and $\delta$.
The qualitative insights presented in this study do not depend on the choice of the link function.

\subsection{Constant recovery rate}

In the simplest reduced--form ansatz  the recovery rate is assumed constant,
\begin{equation}
\langle R(X_\m) \rangle= \langle R \rangle= \mathrm{const} \;.
\end{equation}
Although it is obvious that this crude model neglects important structural information and therefore cannot be accurate, it is still widely used in research and in practice. % CITATIONS !!!!
The only valid statements to make with a constant recovery rate involve setting $\langle R_{k} \rangle$ to zero and thereby estimating worst case losses.

\subsection{Structural recovery rate}

In addition to the reduced--form approach, we consider a functional dependence of recovery rates and default probabilities,
\begin{equation}  \label{eq:RofPD}
\langle R (\PD) \rangle = 
\frac{1}{\PD} \exp\left( -B\, \Phi^{-1}(\PD)+\frac{1}{2}B^2 \right)
\Phi\left( \Phi^{-1}(\PD) - B \right)  \;.
\end{equation}
This result has been derived by \cite{SchaeferKoivusalo2011} in the framework of the Merton model for a correlated diffusion process, but it is also a very good approximation for other processes. %
We call this the structural recovery rate model.
The relation depends only on a single parameter $B$. In the case of the correlated diffusion, $B$ is determined by the parameters of this process.

\section{Model calibration} \label{sec4}

We will use the data obtained in our Monte Carlo simulations for model calibration.
Since we do not employ a model for default probabilities, these will be directly calculated from the number of defaults found in the simulation,
\begin{equation}
\PD(X_{\m}) \approx \frac{N_{\D}(X_{\m})}{K} \;.
\end{equation}
The three different recovery models will be calibrated to the average recovery rates $\langle R(X_\m)\rangle$ found in the simulation.

\subsection{Reduced--form recovery rate}

\begin{figure}[h!]  
\centering
\includegraphics[width=0.49\textwidth]{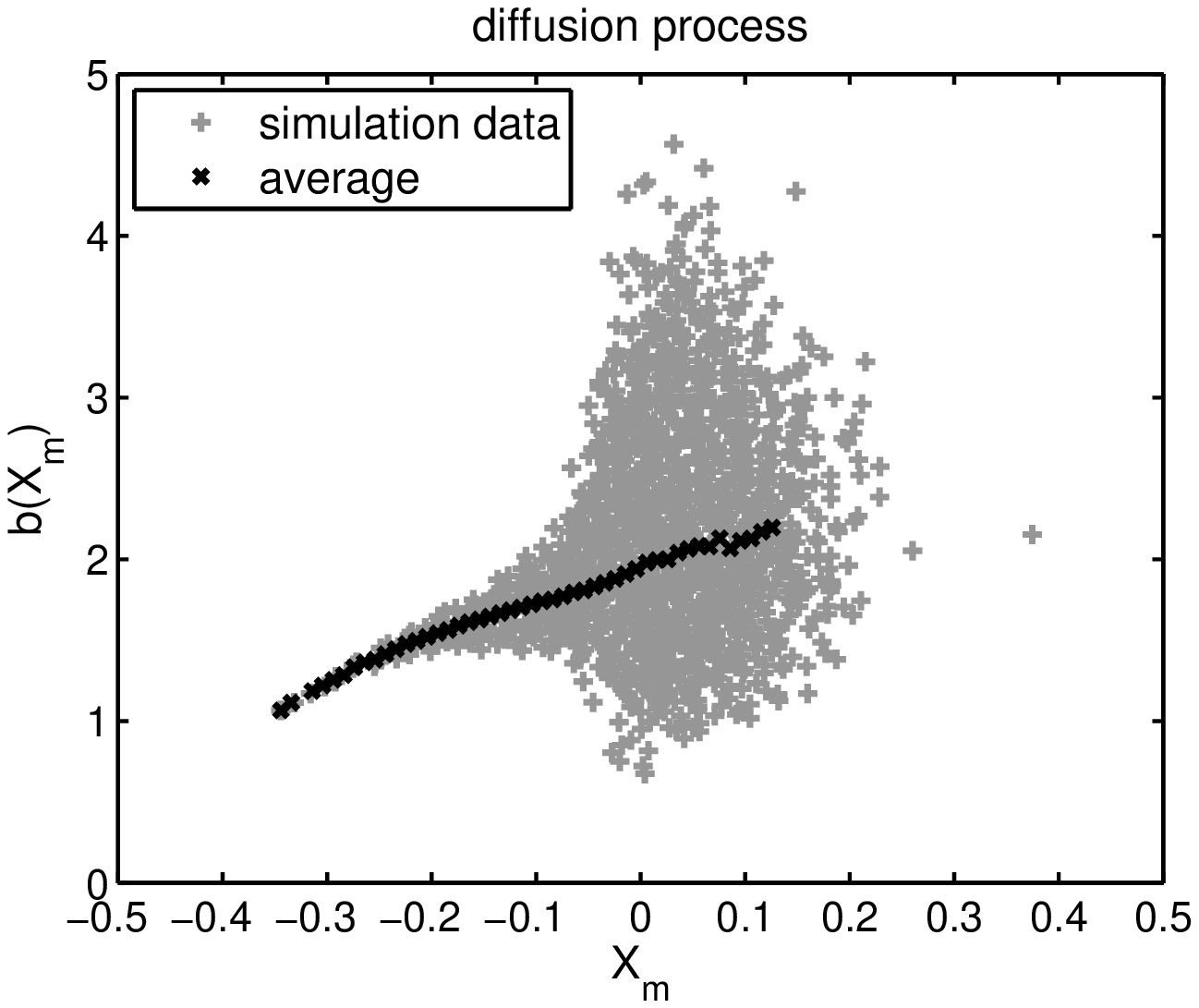}
\includegraphics[width=0.49\textwidth]{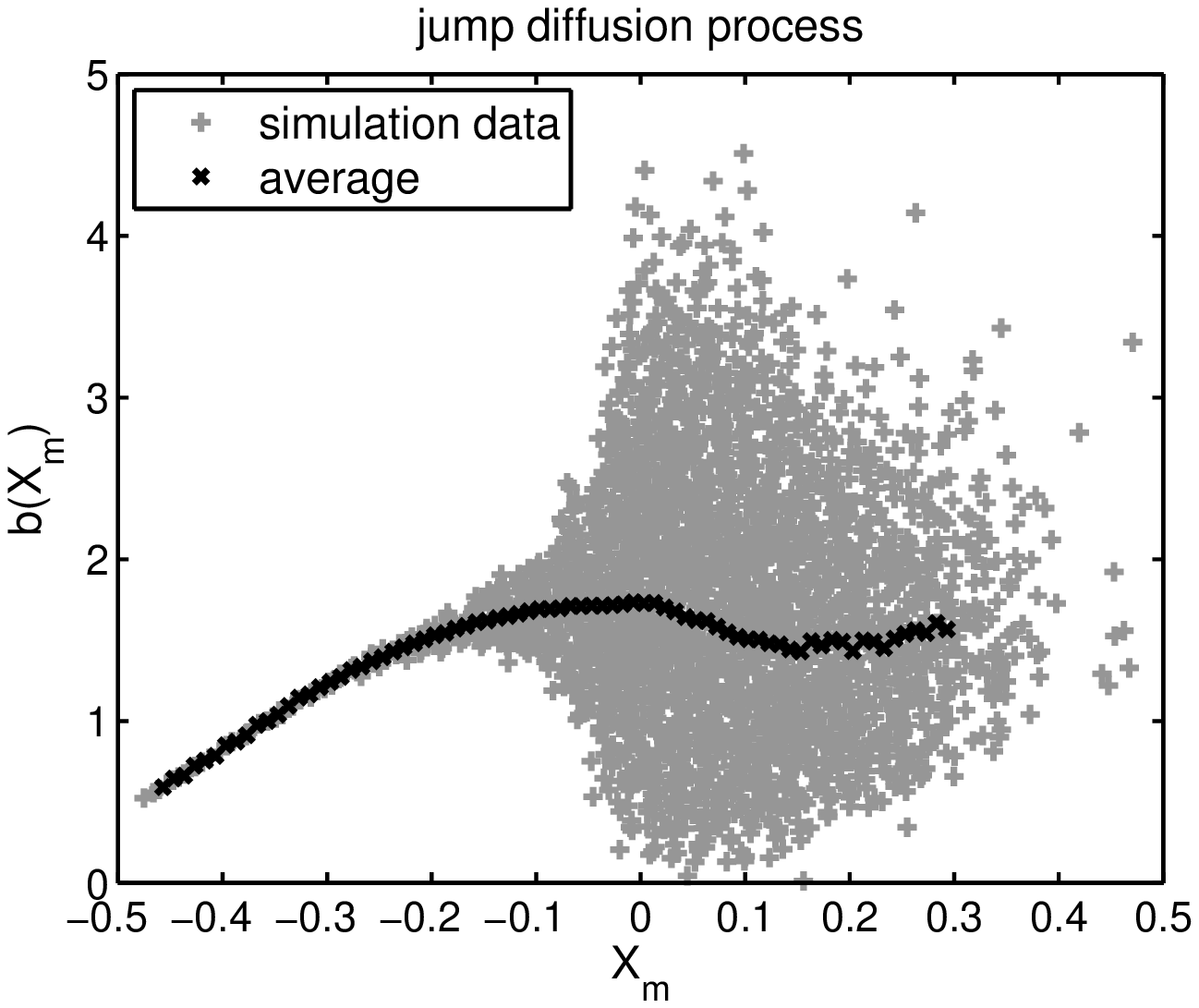}
\caption{
Regression data $b(X_\m)= \Phi^{-1} \left( \langle R(X_{\m}) \rangle \right)$ for the reduced--form recovery model.
The model assumes a linear dependence, which is justified for large negative market returns $X_\m$.
Results are shown for diffusion (left), and for jump--diffusion (right).
} 
\label{fig:b_Xm}
\end{figure}

We apply $\Phi^{-1}$ to both sides of Equation~(\ref{recovery}) and obtain
\begin{equation} \label{inverse_recovery1}
b(X_\m)= \Phi^{-1} \left( \langle R(X_{\m}) \rangle \right) = -\gamma X_\m - \delta  \;.
\end{equation}
According to our reduced--form model we expect $b(X_\m)$ to show a linear dependence.
In Figure~\ref{fig:b_Xm} we present the Monte Carlo data of $b(X_\m)$ for diffusion and jump--diffusion.
In addition to the raw data, we also show average values for small intervals of $X_\m$.
The dependence on $X_\m$ is well defined and rather linear for large negative market returns.
This is true both for diffusion and for jump--diffusion.
For market returns greater than zero, the dependence on $X_\m$ is only visible on average.
In single runs of the simulation, the recovery rates, and thus also $b(X_\m)$,  can deviate considerably from the average values. 
The reason for this is rather obvious: positive market returns make defaults less likely for all individual companies. Thus, defaults and recoveries are mostly determined by the idiosyncratic parts of the processes.
The model assumption of a linear behavior of $b(X_\m)$ is well justified for large negative market returns $X_\m$. However, we observe deviations of this linear behavior starting already around $X_\m=-0.2$. These deviations are most visible in the jump--diffusion case, where we even find a negative slope for $X_\m \ge 0$.
From these findings we can already deduce a qualitative picture of the calibration difficulties that may arise. The overall behavior of recovery rates is not captured by the reduced--form model. If we calibrate only to data for large negative market returns, we will overestimate recovery rates for moderate market returns. And if we only have data available for calibration which corresponds to rather moderate market returns, we would overestimate recovery rates for large negative market returns.
In both cases, we overestimate recovery rates in some situations, and thus underestimate the corresponding losses.
Here we choose 0 as an upper threshold for $X_\m$ and study the calibration results in dependence of a lower threshold. The motivation for this is the typical abundance of data for moderate market returns, while large negative market returns occur only scarcely. We calibrate to the local average values in order to give equal weights to different values of $X_\m$ in the regression.

Another problem becomes obvious when we compare the diffusion and the jump--diffusion case:
The quality of the reduced--form recovery model depends on the underlying process. In other words, a model that works well for one set of credit contracts might not work so well for another.
Non--stationarity in financial time series may also cause the quality of a reduced--form recovery model to degrade over time.

\subsection{Constant recovery rate}

The constant recovery rate is determined as the average value of $\langle R(X_\m)\rangle$ for a given range of market returns $X_\m$. The reasons for choosing an upper and a lower threshold on $X_\m$ have been discussed above.
The upper threshold $X_\m<0$ addresses the bias issue of the estimation at least to some degree.

\subsection{Structural recovery rate}

In our structural recovery approach, we do not fit a covariate dependence of the recovery rate explicitly. Instead, we focus on how recovery rates depend on default probabilities.
The functional dependence of recovery rate on default probability depends only on a single parameter $B$. 
\begin{equation}
\langle R_k (\PD) \rangle = 
\frac{1}{\PD} \exp\left( -B\, \Phi^{-1}(\PD)+\frac{1}{2}B^2 \right)
\Phi\left( \Phi^{-1}(\PD) - B \right)  \;.
\end{equation}
In the Merton model for a correlated diffusion process the parameter $B$ is given by the parameters of this process,
\begin{equation}
B = \sqrt{(1-c)\sigma^2 T} \;.
\end{equation}

\begin{figure}[h!]  
\centering
\includegraphics[width=0.49\textwidth]{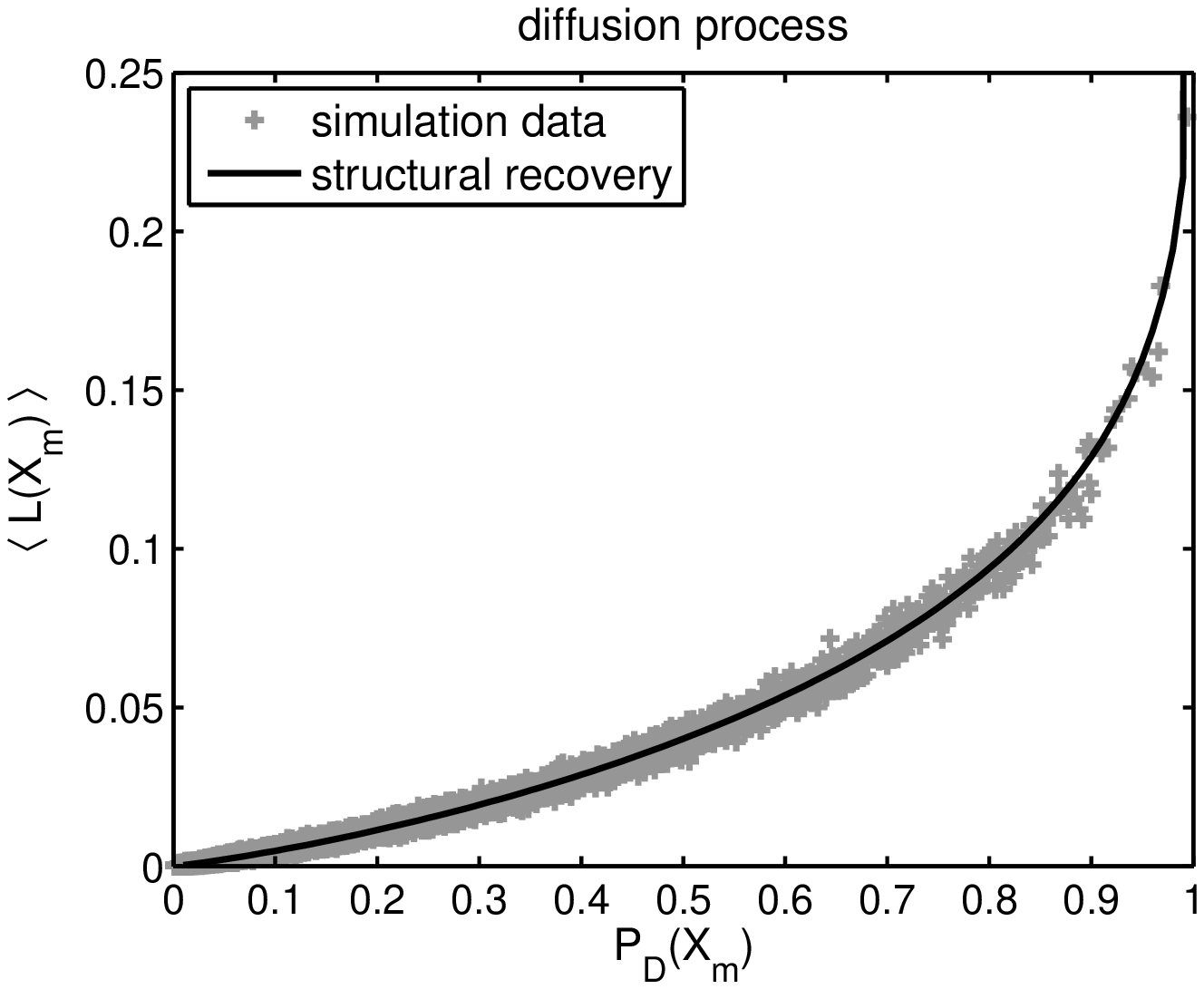}
\includegraphics[width=0.49\textwidth]{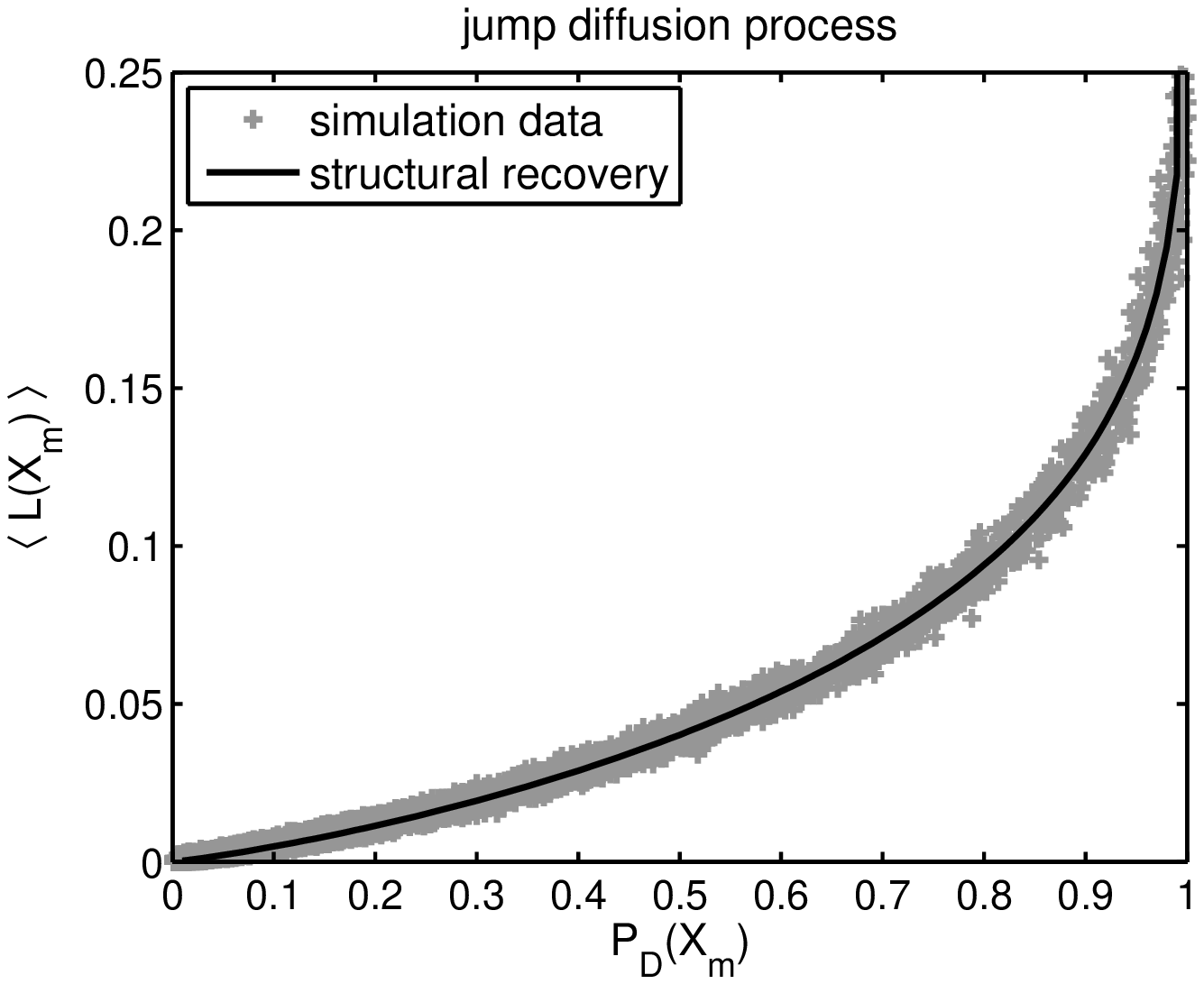}
\caption{
Dependence of average losses $\langle L (X_\m) \rangle$ on default probabilities $\PD(X_\m)$.
Results are shown for diffusion (left), and for jump--diffusion (right).
} 
\label{fig:LofPD}
\end{figure}

We use a least square method to estimate $B$ from the simulation data for $\PD(X_\m)$ and $\langle R_k (X_\m) \rangle$. 
Again, the data are filtered with respect to  an upper and a lower threshold on the market return $X_\m$. The convergence of the regression for $B$ is strong. 
As already pointed out in~\cite{SchaeferKoivusalo2011}, this functional dependence does not critically depend on the underlying process. In fact, it has been shown to work well for diffusion, jump diffusion and GARCH(1,1). Therefore we do not expect a critical dependence of the calibration results on the range for $X_\m$.
We can further improve the stability of the calibration if we regress the data of average losses directly. Inserting the functional dependence (\ref{eq:RofPD}) into Equation (\ref{eq:LofPD1}) yields
\begin{equation}
\langle L (\PD) \rangle = 
\PD - \exp\left( -B\, \Phi^{-1}(\PD)+\frac{1}{2}B^2 \right)
\Phi\left( \Phi^{-1}(\PD) - B \right)  \;.
\end{equation}
Now the strongly fluctuating results for small default probabilities are suppressed in the regression, see Figure~\ref{fig:LofPD}. This leads to an improvement in calibration stability.

\section{Comparison of model results} \label{sec5}

\subsection{Dependence of recovery rates and default probabilities}
\begin{figure}[h!] 
\centering
\includegraphics[width=0.49\textwidth]{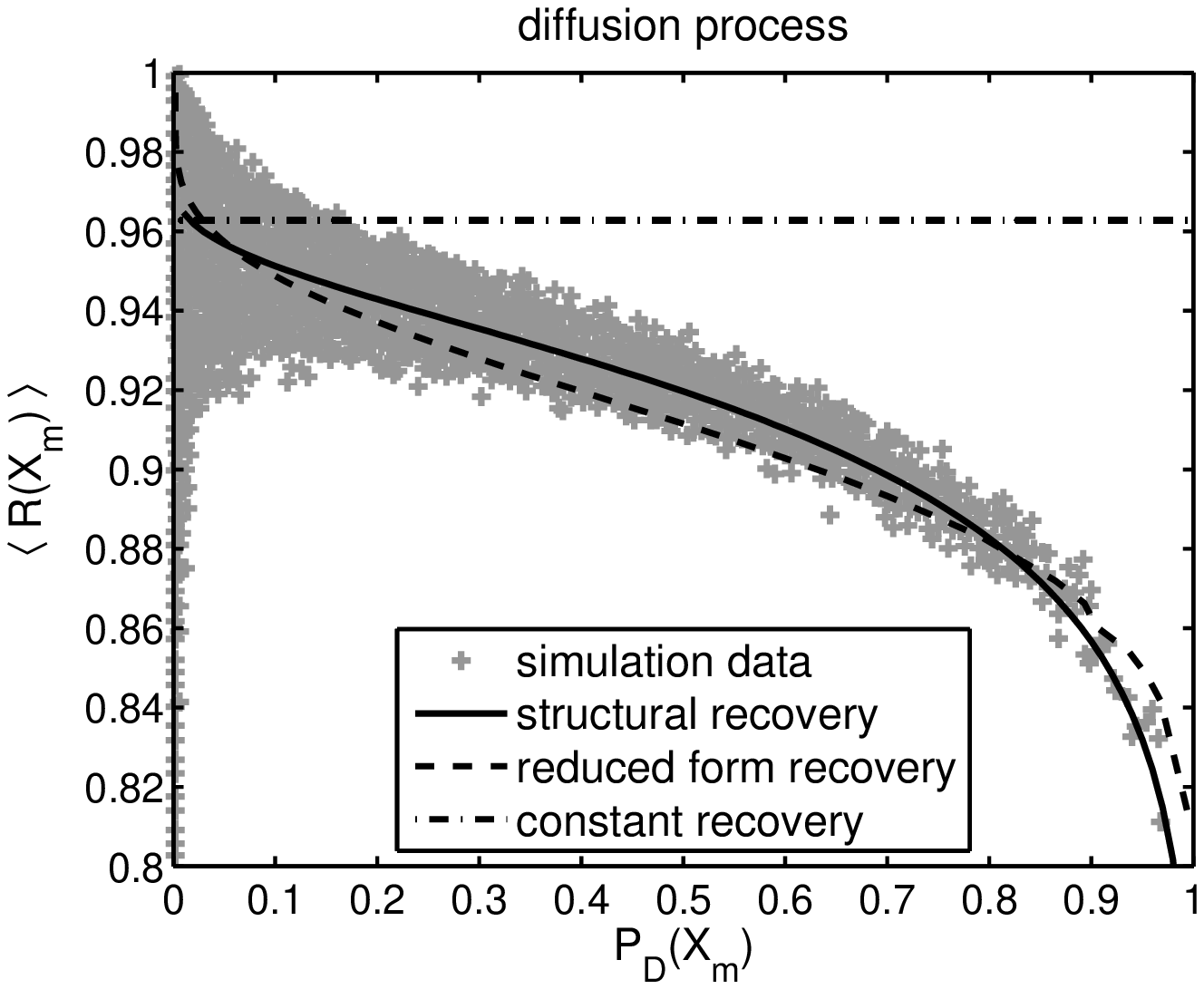}
\includegraphics[width=0.49\textwidth]{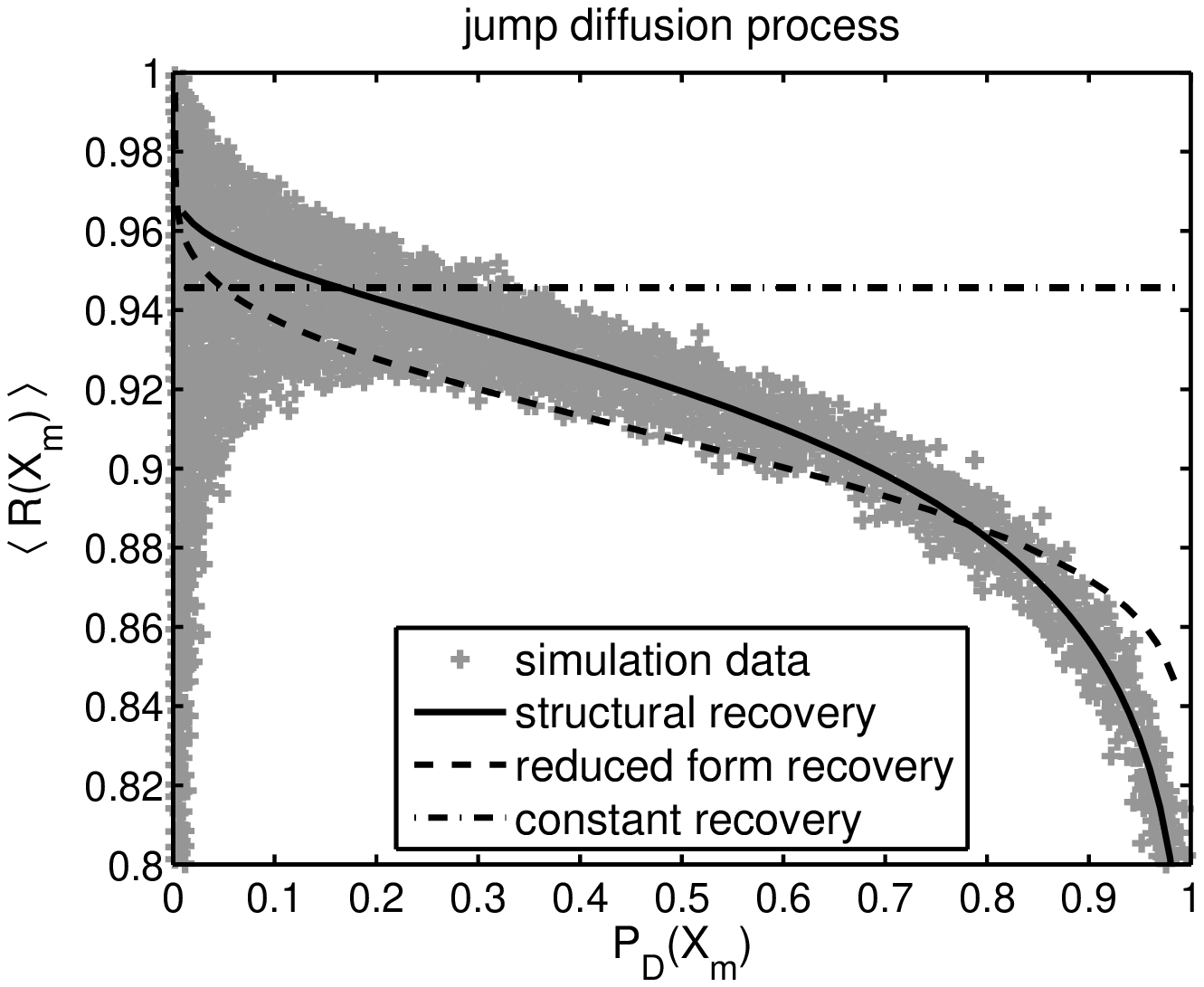}
\caption{
Dependence of recovery rates $\langle R(X_\m) \rangle$ on default probability $\PD(X_\m)$. 
The Monte Carlo results are compared to structural (solid line), reduced--form (dashed line) and constant recovery rate (dashed-dotted line).
Results are shown for diffusion (left), and for jump--diffusion (right).
}
\label{fig:R_PD}
\end{figure}

Let us first turn our attention to the dependence between recovery rates and default probabilities.
Figure \ref{fig:R_PD} shows scatter plots of the Monte Carlo results for $\PD(X_\m)$ and $\langle R(X_\m) \rangle$. We compare these with the results for the three different recovery rate models, where  $\PD(X_\m)$ is again taken directly from the simulation.
Here, the model parameters for the recovery rate are calibrated to the data with $X_\m < 0$.  
It is obvious that the assumption of a constant recovery rate is not justified at all.
Moreover, the estimation of this constant is biased due to the relative abundance of data for small default probability and high recovery rate. Although this bias could be taken into account, the results for the constant recovery rate model appear arbitrary and neglect important structural information.
For the reduced--form recovery, we notice that the interdependence of recovery and default rates is preserved at least qualitatively. We observe larger deviations for the jump--diffusion. This is to be expected, since the recovery model does not describe the simulation results over the whole data range, see our discussion in the previous section.
The qualitative agreement of the reduced--form model results and the simulation data can be attributed to the fact that both $\PD(X_\m)$ and $\langle R(X_\m) \rangle$ describe the dependence on the same covariate. This is in accordance with findings of \cite{Chava2008} who pointed out that reduced--form models can yield considerably better results, if defaults and recoveries are modelled and calibrated with respect to the same set of covariates.

The structural recovery describes the data very well over the whole data range. 
This is not surprising for the diffusion process, after all the functional dependence has been derived for this case. It is noteworthy, however, that we find such a perfect agreement for the jump--diffusion as well. This result was previously mentioned in \cite{SchaeferKoivusalo2011}.

\subsection{Value at Risk and Expected Tail Loss}  

Finally, we want to test the quality of the model results on two risk measures that capture the tail behavior of the loss distribution. For this purpose, we consider the Value at Risk and the Expected Tail Loss, or Expected Shortfall. Both measures are commonly used in credit risk manangement, see, e.g., \cite{Artzner1997,FreyMcNeil2002,YamaiYoshiba2005}, and in financial regulations, see \cite{Basel2}. 
For details on the calculation of these risk measures we refer to \ref{sec:riskmeasures}.
As mentioned above, we are considering a situation where the default probabilities are well described by a model --- or in our case by the simulation itself. Hence, we focus entirely on the influence of the different recovery rate models.
In particular, we are interested in the impact of limited calibration data on the model results.
To this end we filter the simulation data with respect to the market return $X_\m$.
As before we keep 0 as an upper threshold, because default and recovery rates do not show a clear dependence on the market return for positive $X_\m$.
Additionally, we set a lower threshold on $X_\m$ and study the results for VaR and ETL in dependence of this lower threshold. 
This corresponds to the relative abundance of empirical data for moderate market returns, while large negative market returns occur only scarcely. Frankly, we ask to which extend the different models can describe extreme events when they are only calibrated with default and recovery data from a relatively calm period.

\begin{figure}[h!]
\centering
\includegraphics[width=0.49\textwidth]{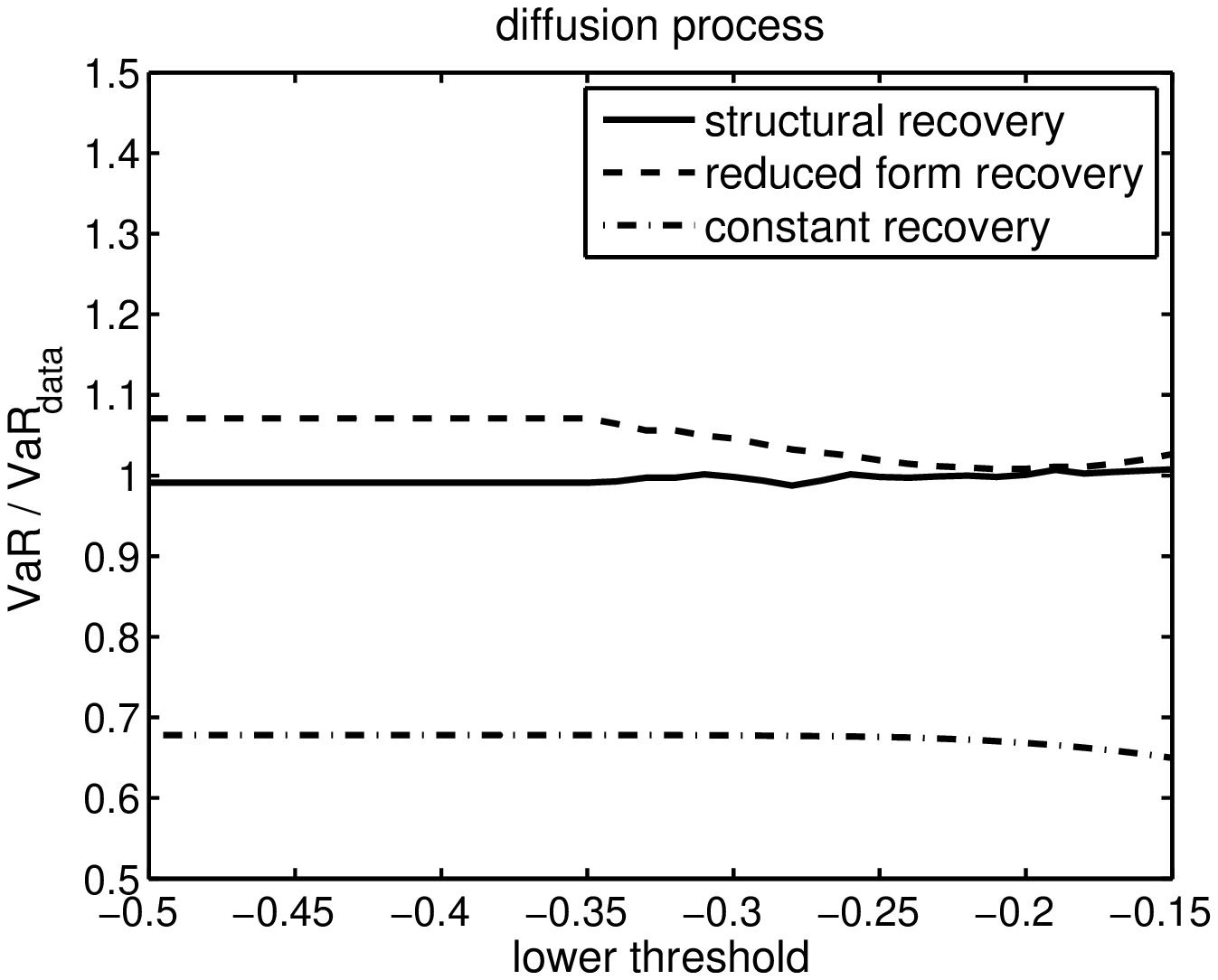}
\includegraphics[width=0.49\textwidth]{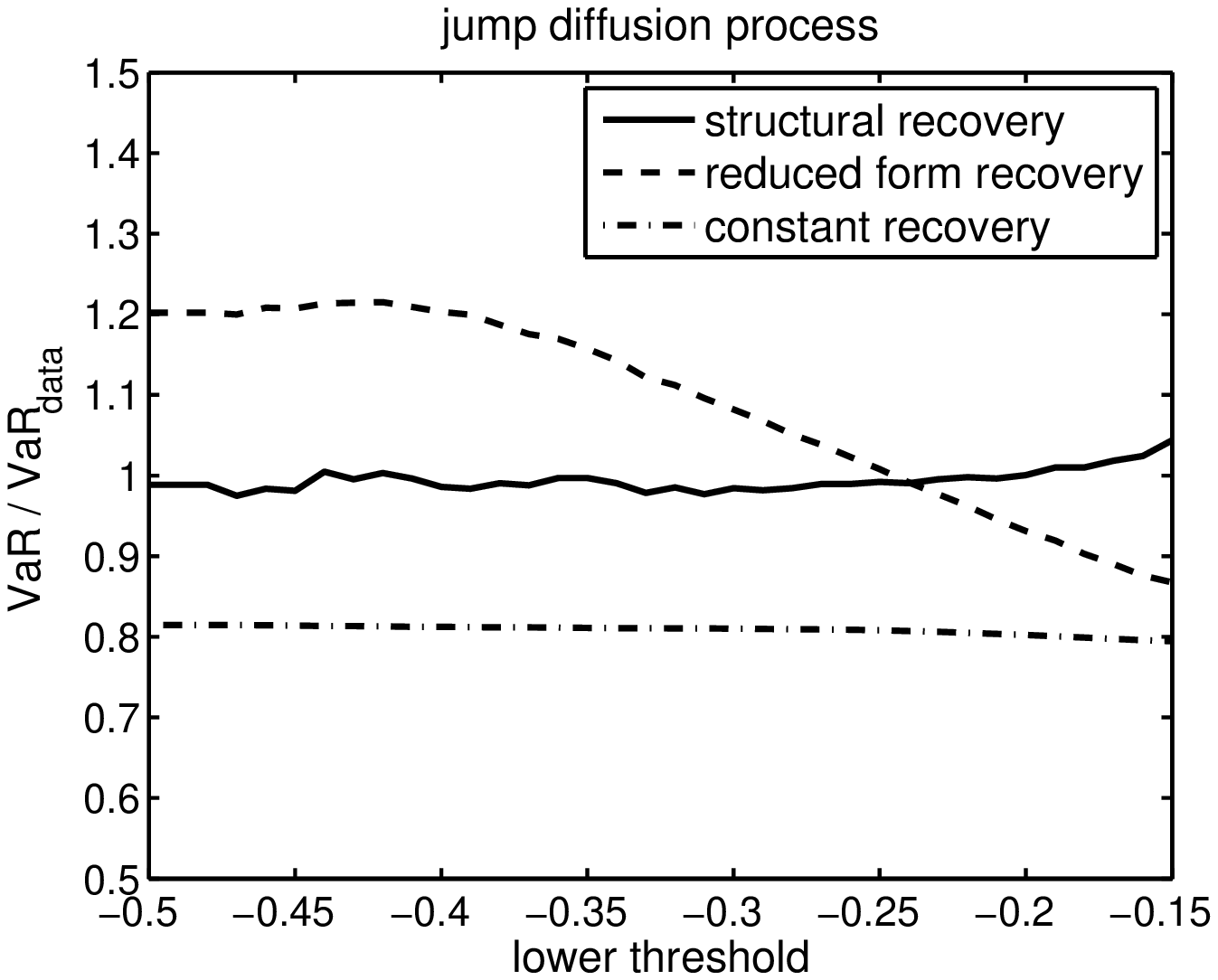}
\caption{
Value at Risk ${\rm VaR}_\alpha$ at the confidence level $\alpha=0.99$. The results are normalized to the Value at Risk computed for the entire data set of the simulations. Thus, values above 1 correspond to an overestimation, values below 1 to an underestimation by the model.
Results are shown for diffusion (left), and for jump--diffusion (right).
}  
\label{fig:VaR}
\end{figure}

Figure \ref{fig:VaR} shows the model results for the Value at Risk ${\rm VaR}_\alpha$ at the confidence level $\alpha=0.99$. The results are normalized to the corresponding empirical value which is computed for the entire data set of the simulation. 
These empirical values are ${\rm VaR}_{0.99}=0.013$ for the diffusion and ${\rm VaR}_{0.99}=0.015$ for the jump--diffusion.
Not surprisingly, we find the worst results for the constant recovery. The Value at Risk is underestimated by more than 30\% in the diffusion case and by almost 20\% in the jump--diffusion case.
The better result for the jump--diffusion is misleading, however, as we will see later on.
The reduced--form recovery yields quite reasonable results in the diffusion case. It overestimates the Value at Risk by roughly 8\% when all data with $X_\m<0$ is taken into account. We see a slight dependence on the lower threshold for $X_\m$. This starts only at $X_\m\approx-0.35$, because this was the lowest market return observed in the simulation. 
In the jump--diffusion case, the reduced--form model yields very unstable results, ranging from a considerable overestimation to a severe underestimation of the Value at Risk.
For both processes, the structural recovery model provides rather stable results with very little deviation from the empirical values.

\begin{figure}[h!]
\centering
\includegraphics[width=0.49\textwidth]{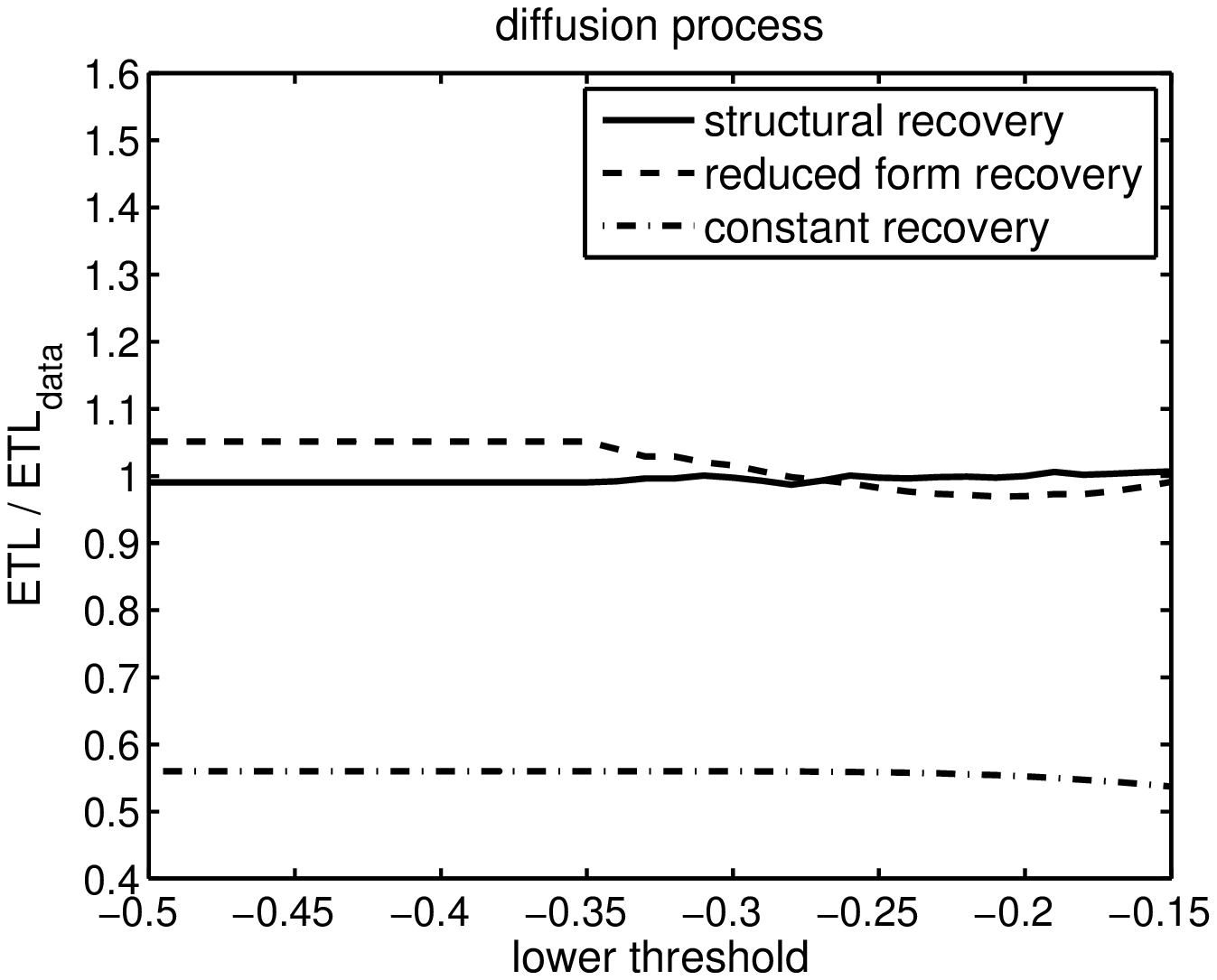}
\includegraphics[width=0.49\textwidth]{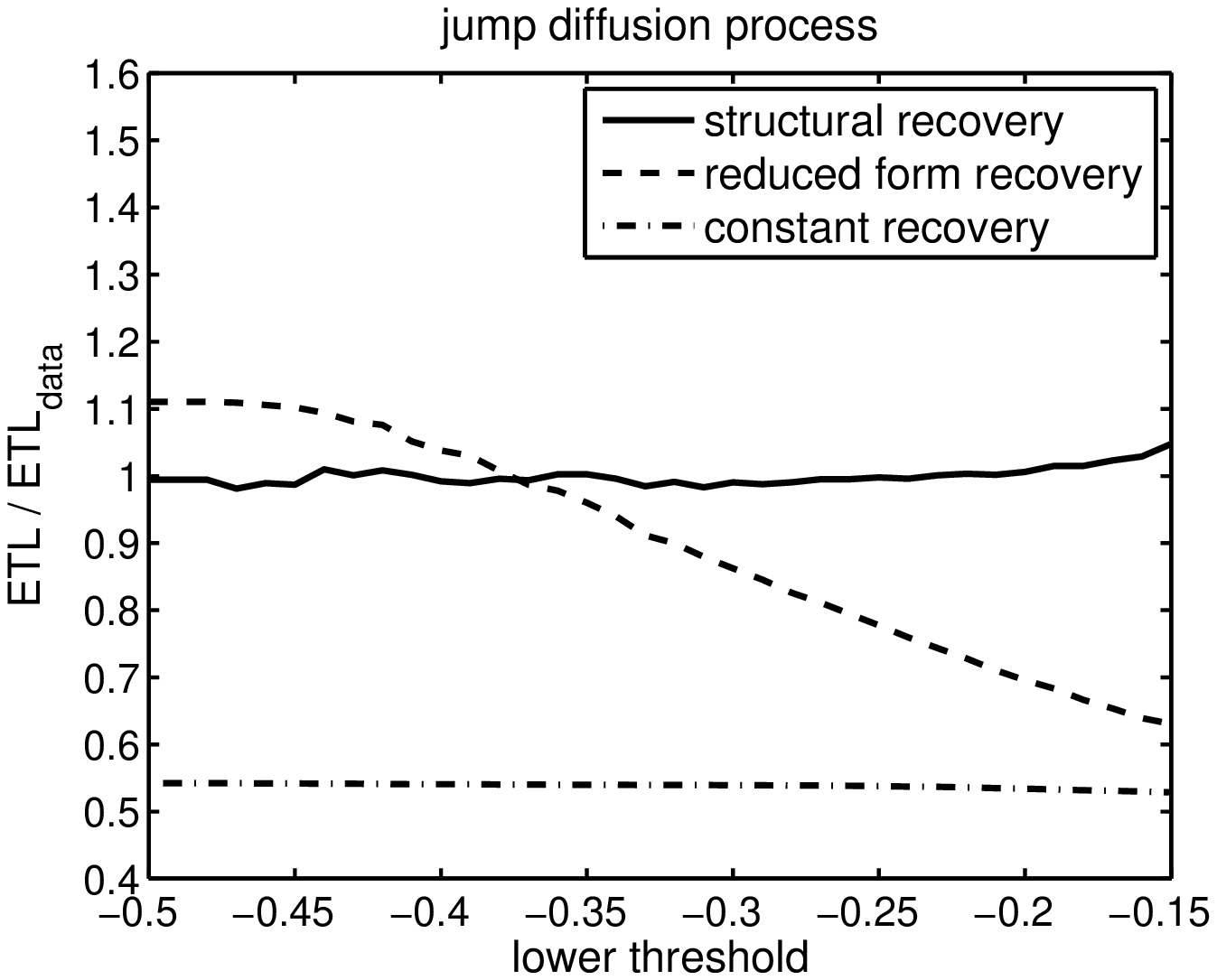}
\caption{
Expected Tail Loss ${\rm ETL}_\alpha$ at the confidence level $\alpha=0.99$. The results are normalized to the ETL computed for the entire data set of the simulations. Thus, values above 1 correspond to an overestimation, values below 1 to an underestimation by the model.
Results are shown for diffusion (left), and for jump--diffusion (right).
}   
\label{fig:ETL}
\end{figure}

In Figure \ref{fig:ETL} we present the model results for the Expected Tail Loss ${\rm ETL}_\alpha$  at the confidence level $\alpha=0.99$. Again, we normalized the results to the empirical value for the entire data set of the simulation. 
These empirical values are ${\rm ETL}_{0.99}=0.0238$ for the diffusion and ${\rm ETL}_{0.99}=0.0379$ for the jump--diffusion.
The Expected Tail Loss is better suited to describe the tail behavior of the loss distibution, as first pointed out by \cite{Artzner1997}.
Here we see that the constant recovery rate does in fact not provide a better description of the jump--diffusion than of the diffusion.
The results for the Value at Risk are more coincidental, they depend critically on the chosen confidence level $\alpha$.
The reduced--form recovery again shows rather good results for the diffusion process. However, the Expected Tail Loss reveals even more clearly that the reduced--form recovery severly underestimates large losses, if the calibration data is restricted. Moreover, we observe a very strong dependence of the results on the lower threshold for the market returns. 
These unstable results are contrasted by the stable results for the structural recovery rate. The Expected Tail Loss confirms that this model delivers a very good and robust estimate for large portfolio losses.

\section{Conclusions} \label{sec6}

The relation between recovery rates and default probabilities has a pronounced effect on the tail of the portfolio loss distribution. Yet both quantities are often modelled independently in current credit risk models.
In this paper we consider a Merton model as a reference system with diffusion and jump--diffusion as underlying processes.
Three different recovery models are calibrated to Monte Carlo simulations of this reference system: a reduced-form recovery model, the special case of constant recoveries, and the structural recovery model. A functional dependence between default and recovery rates can be derived for the Merton model with correlated diffusion. The structural recovery model employs this same functional dependence regardless of the underlying process.
We study to what extend the different recovery models preserve the relation between default and recovery rates. Most importantly, we compare how well they reproduce two measures of tail risk, the Value at Risk and the Expected Tail Loss.
Furthermore, we discuss calibration issues that may arise due to a lack of information. To this end we confine the calibration to a certain range of the simulation data. A lower threshold on the market return implies that extremely adverse market situations are not contained in the historical data sample on defaults and recoveries.

It comes as no surprise that the constant recovery rate model shows the worst performance, 
since it does not include a dependence on covariates or default probabilities.
In addition, the estimate of the average recovery rate from historical or simulation data is biased due to the relative abundance of small losses.
As a consequence this model shows very poor results for Value at Risk and Expected Tail Loss.
Nonetheless we included this model in our discussion, as it is still prevalent in practical applications.

The results for the reduced--form recovery model are considerably better.
The generalized linear dependence on market returns works reasonably well, especially when considering large negative market returns.
However,  it does not provide a good description over entire data range.
Furthermore, the quality of the model critically depends on the underlying process.
The model is able to reproduce the dependence of default and recovery rates at least qualitatively. Larger deviations are observed for the jump diffusion process.
We find a critical dependence of the model results on the data range used for calibration.
This leads to very unstable estimates for Value at Risk and Expected Tail Loss.

The structural recovery model accurately describes the dependence of default and recovery rates for different processes.
It has only a single parameter and is calibrated with respect to the default rate dependence.
The model yields nearly perfect estimates of Value at Risk and Expected Tail Loss.
These results show almost no dependence on the data range used for calibration.

In accordance with \cite{Chava2008} we conclude that a reduced--form ansatz can work well, if the same covariates are used for both default and recover rates.
However, an additional model risk is involved in the reduced--form recovery with respect to the underlying process.
The structural recovery avoids this model risk because it does not rely on covariates, instead it directly describes the dependence of recovery rates on default probabilities. Thus, it can be easily used with any default model.

\section*{Acknowledgments}
We wish to thank Thomas Guhr and Sven {\AA}berg for helpful discussions.

\appendix
%%%
\section{Calculation of risk measures}  \label{sec:riskmeasures}  
In the reduced--form approach default probabilities and recovery rates are modelled in dependence of covariates, in our case, in dependence of the market return $X_{\m}$. The average loss or loss of a homogeneous portfolio is then 
\begin{equation} \label{eq:plossrec1}
\left< L(X_{\rm m}) \right> = \PD(X_{\rm m}) \left( 1-\left<R(X_{\rm m})\right> \right) \;.
\end{equation}
There is an abundance of empirical data on covariates, we can therefore assume that their distribution is well understood and modelled. In the case of a correlated diffusion, we know that the market return is log-normal distributed. The jump--diffusion process renders the tail of the distribution fatter, corresponding more closely to the situation on the stock market.
When the PDF $p_{X_\m}(X_\m)$ is known, 
we can use the substitution
\begin{equation}
\left| p_L(L) \rmd L \right| = \left| p_{X_\m}(X_\m) \rmd X_\m \right|
\end{equation}
to calculate the Value at Risk and the Expected Tail Loss.
Here we simplified the notation to $L$ instead of $\left< L(X_{\rm m}) \right>$.
With the cumulative distribution function
\begin{align}
F_{X_{\m}}(x)=\int\limits_{-\infty}^{x} p_{X_{\m}}(X_{\m}) \rmd X_{\m}
\end{align}
we can express the Value at Risk $\mathrm{VaR}_{\alpha}$ as
\begin{align}
\mathrm{VaR}_{\alpha}=L(F_{X_{\m}}^{-1}(1-\alpha)) \;,
\end{align}
where the function $L$ is the $X_\m$-dependence given in Equation~(\ref{eq:plossrec1}).
The value $F_{X_{\m}}^{-1}(1-\alpha)$ is the $1-\alpha$ quantile of the market return. For numerical data it can be calculated directly.

The Expected Tail Loss is calculated as
\begin{equation}
{\rm ETL}_\alpha = \frac{1}{\alpha} \int\limits_{{\rm VaR}_\alpha}^1 L\, p_L(L) \, \rmd L  
= \frac{1}{\alpha} \int\limits_{-\infty}^{F_{X_\m}^{-1}(1-\alpha)} L(X_\m)\, p_{X_\m}(X_\m) \, \rmd X_\m  \;.
\end{equation}
For numerical data we can determine the Expected Tail Loss as
\begin{equation}
{\rm ETL}_\alpha = {\rm mean} \left\{  L(X_\m) \; \big| \; X_\m \le F_{X_\m}^{-1}(1-\alpha) \right\} \;.
\end{equation}
The same line of reasoning applies for the structural recovery approach. The average loss is then given by
\begin{equation}\label{eq:LofPD1}
\left< L(\PD) \right> = \PD \left( 1-\left<R(\PD)\right> \right) \;,
\end{equation}
with the expected recovery rate according to Equation~(\ref{eq:RofPD}).
When we know the distribution of default probabilities, e.g., by means of modelling, by a scenario analysis or from Monte Carlo simulations, we can express the Value at Risk as
\begin{align}
\mathrm{VaR}_{\alpha}=L(F_{\PD}^{-1}(\alpha)) \;,
\end{align}
where $F_{\PD}$ is the cumulative distribution function of the default probability.
For the Expected Tail Loss we find
\begin{equation}
{\rm ETL}_\alpha  
= \frac{1}{\alpha} \int\limits_{F_{\PD}^{-1}(\alpha)}^1 L(\PD)\, p_{\PD}(\PD) \, \rmd \PD  \;.
\end{equation}
For numerical data we can evaluate it as
\begin{equation}
{\rm ETL}_\alpha = {\rm mean} \left\{  L(\PD) \; \big| \; \PD \ge F_{\PD}^{-1}(\alpha) \right\} \;.
\end{equation}

%\section*{References}
%\bibliographystyle{plainnat}
\bibliographystyle{elsarticle-harv}
\bibliography{econophysics,creditrisk}

\end{document}